\documentclass[aps,prl,reprint,superscriptaddress]{revtex4-2}
\usepackage{amsmath,amssymb,bm}
\usepackage{microtype}
\usepackage[colorlinks=true,allcolors=blue]{hyperref}
\usepackage{graphicx}
\usepackage{amsfonts}
\usepackage{comment}
\usepackage{verbatim}
\usepackage{mathtools}
\usepackage{refcount}

\makeatletter
\newcommand\footnoteref[1]{\protected@xdef\@thefnmark{\ref{#1}}\@footnotemark}
\makeatother

\begin{document}

\title{Optimal Search Strategies under Finite Resource Constraints}

\author{Hugues Meyer}
\affiliation{School of Physics and Astronomy, University of Nottingham, Nottingham NG7 2RD, United Kingdom}
\affiliation{Centre for the Mathematics and Theoretical Physics of Quantum Non-Equilibrium Systems, University of Nottingham, Nottingham NG7 2RD, United Kingdom}

\date{\today}

\begin{abstract}
Improving target search is a challenge faced by many natural and artificial agents, often requiring the use of energetic or material resources. These inputs may be supplied by an external controller or generated by the searcher itself, for example, through the deposition of cues along its trajectory. Although such resources are commonly modeled as costly but available without an explicit limit, in practice, they can often only be produced in finite amounts. Here, we go beyond the assumption of an unlimited supply of resources by investigating searches in which a finite resource budget must be allocated throughout the search process. To investigate this problem, we equip a one-dimensional diffusive searcher with three types of resources: an external force, a supply of partially reflecting flags, and auto-chemotactic cues. We derive exact optimal resource-allocation protocols for informed search processes in which the agent has continuous access to its past and current positions, which we find to always require resource allocation to be spatially concentrated.  We then determine the corresponding protocols for blind searchers and show that resource usage must be more evenly distributed. Non-Markovian cues that generate environmental memory are found to substantially improve search performance when the agent has access to positional information, even under finite-resource constraints. In contrast, their benefits are considerably more limited for blind search, particularly when the cues provide no directional information.
\end{abstract}

\maketitle

Efficient target search is a universal problem encountered across a broad range of  natural and artificial processes at various scales \cite{Grebenkov2024}, including chemical reaction kinetics \cite{Smoluchowski1918, Collins1949, Rice1985}, genetic transcription \cite{Berg1981, Gorman2008}, immune responses \cite{DeOliveira2016, Krummel2016, Levin2016, Fricke2016}, animal and human foraging \cite{Hassell1978, Traniello1989, Hills2013}, and search-and-rescue missions \cite{Drew2021, Lyu2023, Yedilkhan2025}. Search optimization has become a central topic in statistical physics, with various mechanisms identified to improve efficiency, including Lévy walks \cite{Raposo2003, Edwards2007}, intermittent strategies \cite{Lomholt2008,  Benichou2011, Bressloff2020a}, stochastic resetting \cite{Chechkin2018, Wald2021, Olsen2026, Evans2020, Gupta2022, Pal2024}, persistent motion \cite{Tejedor2012}, collective searches \cite{MejiaMonasterio2011, Meyer2024, Grebenkov2025}, trail-leaving walks \cite{Wysocki2026, Pineau2026}, non-Markovian dynamics \cite{Meyer2021, Barbier-Chebbah2024, Coghi2025, Dolgushev2025, Bremont2025, Netz2026b}, and activity \cite{Rupprecht2016, Basu2024, Jeon2024}. Such strategies often require additional energy or resources to modify the search dynamics, such that a  trade-off between search efficiency and the costs associated to the implementation of the strategy must often be found, for instance in resetting protocols \cite{Sunil2023, Tal-Friedman2025, Biswas2025}, collective searches \cite{Meyer2025}, or optimal control \cite{Davis2026}. However, resource availability is generally assumed to be unlimited, in particular in non-Markovian processes where searchers exploit internal or external resources to leave cues of their passage and generate environmental memory along their trajectories \cite{Hokmabad2022, Veres2022, Nakayama2023,Ipina2024, Altshuler2024, Singh2026}.
Here, we investigate search processes under a finite resource budget, which cannot be replenished during the search. Rather than minimizing a weighted combination of search time and resource expenditure, we seek to allocate a fixed budget to minimize the typical search time. This setting reflects realistic situations in which resource availability is predetermined \cite{Wolf1975, Ydenberg1994, Giang2024}, yet has received comparatively little attention \cite{Sato2010, Grebenkov2022, Linn2026}.
The central challenge is to exploit the available budget without exhausting it too early or leaving resources unused when the target is reached. 

To tackle this finite resource allocation problem, our reference resource-free process is a single one-dimensional diffusive walker initialized at $x=0$ on the semi-infinite line, which aims to reach a target located at position $x=L$ \footnote{Note that we consider a fixed target whose location is known, although generalizing to a randomly distributed target within an interval does not alter the main claims and phenomenology of our results, as discussed at the end of this letter}. The mean first-passage time (MFPT) for such a process is known to be $\bar{\mathcal{T}}_0 =L^2/2D$  where $D$ is the diffusion constant of the searcher \cite{Redner2001}. Providing the searcher with a finite budget of resources usable along its path may allow to reduce this reference search time. Such resources can be actionable at a distance by an external agent, or carried by the searcher itself to be used throughout its trajectory. The latter case implies the process to be self-interacting and {\it a priori} non-Markovian. Moreover, the resource may or may not carry information about the location of the target and help to direct the searcher towards it. We therefore consider three resource models that differ in whether resources are deposited by the searcher itself and whether they provide directional information that can guide the searcher towards the target.

In model I, an external controller applies an external force $F$ to the searcher that pushes it towards the target, which constitutes our memory-free model with directional information. The resource usage is fixed by a finite work budget, that is the total work performed by the force  -- or equivalently the total heat dissipated --  until the searcher reaches the target must never exceed a prescribed value $W$ at any point along the trajectory. 

In model II, the resource still carries information about the target location but introduces environmental memory. In fact, we equip the searcher with directional {\it flags} against one side of which it can be reflected with probability $p$. In the lattice version of the model, the searcher may deposit at each time step a flag taken from a finite supply of $N$ flags on the lattice bond between the site $i_n$ it currently occupies and the site $i_n-1$. The number of flags on that bond if noted $f_{i_n}$. Then, it can hop to a neighboring site $i_{n}\pm 1$ with probabilities
\begin{gather}
    p_{i_n\to i_n-1} = \frac{(1-p)^{f_{i_n}}}{2}, \qquad p_{i\to i+1} = \frac{1}{2},
\end{gather}
or remain on its current site with probability $ p_{i_n\to i_n} = 1 - \sum_{j=\pm1} p_{i_n\to i_n+j}$. The presence of flags therefore only hinders backward transitions. In the continuous limit, flags act as as potential drops of height $\beta^{-1}|\ln(1-p)|$, hence generating a piecewise constant potential profile.

Model III still includes environmental memory but the resource loses directional information. Adapting the well-known {\it true} self-avoiding walk \cite{Amit1983}, the searcher walks on a lattice over the sites of which a cue field $c_i$ is defined. At each time step, the walker can deposit an amount $\delta c_n$ on its current site $i_n$ which it draws from a supply whose initial size is finite and equal to $R_0$. Then, it hops from its current site $i_n$ to one of the neighboring sites $i_n \pm 1$ with probability 
\begin{gather}
    \label{eq:TSAW}
    p_{i_n\to i_n\pm1} = \left[ 1 + e^{\pm \beta \left(c_{i_n + 1}- c_{i_n - 1}\right)}\right]^{-1}
\end{gather}
where $\beta>0$ acts as an effective inverse temperature or repulsive interaction strength.

\begin{figure}
    \centering
    \includegraphics[width=\linewidth]{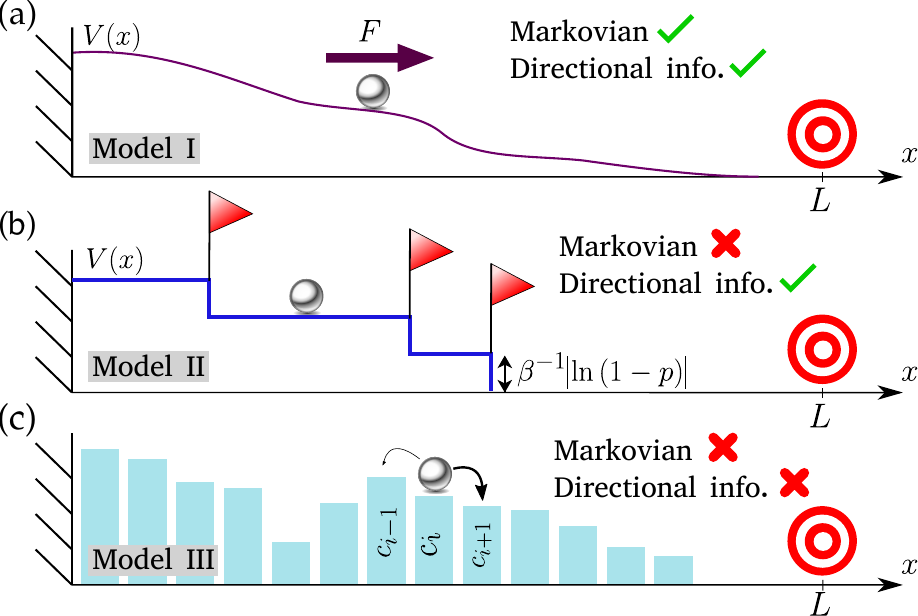}
    \caption{Sketch of the models, in which the searcher starts at $x=0$ and needs to reach $x=L$ on the semi-infinite line. (a) An external force is applied via a potential profile. (b) Flags are deposited by the searcher along its way, which partially reflect it and effectively act as a piecewise constant potential. (c) Continuous cues can be deposited by the searcher and its further motion is governed by the cue profile gradient. }
    \label{fig:models}
\end{figure}

The three models are sketched in figure \ref{fig:models}. They all allow the searcher a certain freedom to decide on a strategy for how to use the finite resource budget to reach the target as fast as possible. Approaches to tackle this decision-making problem may significantly vary depending on the information available about the searcher's position, which leads to consider two classes of strategies. In {\it informed} protocols, the searcher knows at all times its current position -- although it may not be able to track time -- and can therefore design a {\it spatial} protocol that determines whether the resource is used or not based on the searcher's past and current positions. Conversely, in {\it blind} protocols, the searcher has no information on its current and past locations and can only follow a predetermined temporal protocol for how to use the resource. The challenge is then to maximize the probability of exhausting the budget upon reaching the target without spending it too fast and leave most of the search without available resource.\\

\begin{figure*}
    \includegraphics[width = \linewidth]{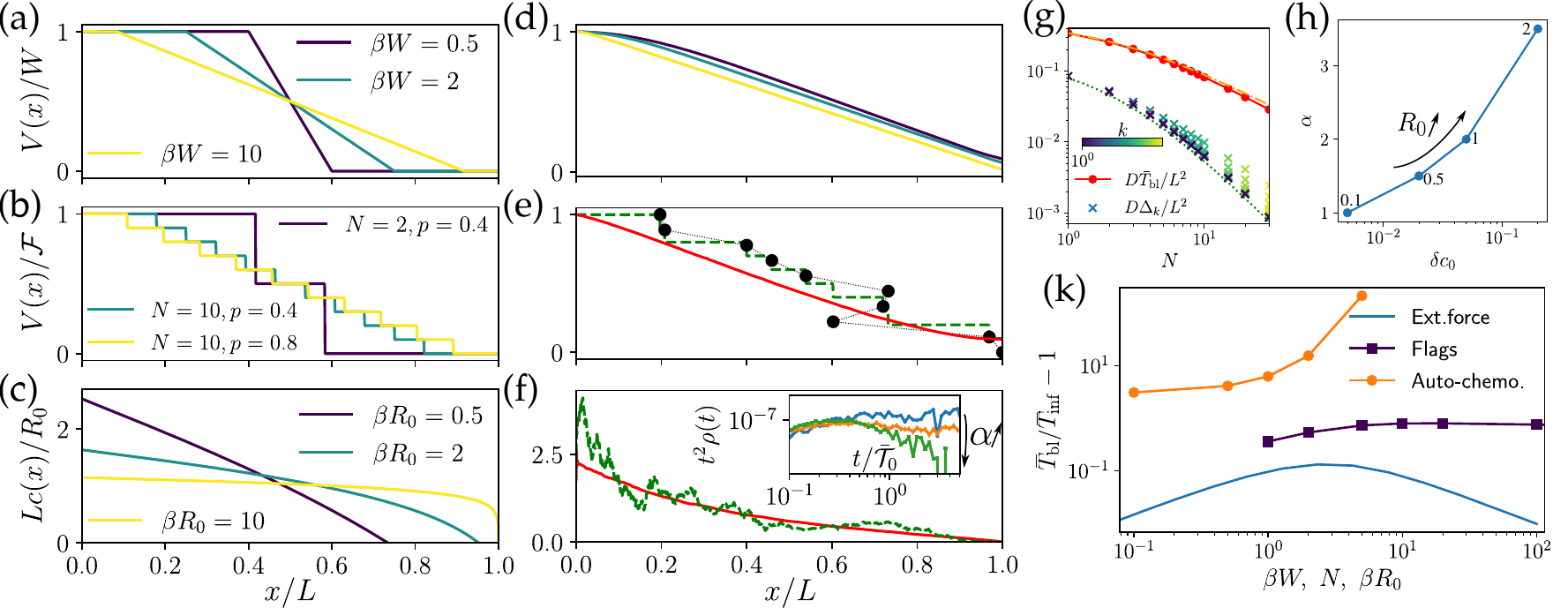}
    \caption{(a-c) Optimal resource profile in the informed processes: (a) potential profile in model I, (b) flag profile in model II (c) auto-chemotactic cue profile in model III. (d-e) Trajectory-averaged resource profiles in the optimal blind processes: (d) Trajectory-averaged potential in model I; (e) Trajectory-averaged potential generated by deposited flags in model II for $N=10$ and $p=0.8$ (solid line), together with a profile of a single trajectory (dashed line) where the dotted line and dots show the order in which flags were deposited. (f) Trajectory-averaged chemotactic cue profile for $\beta R_0/L = 2$ (solid line) with one instance of a profile in a single trajectory (dashed line). The inset shows $t^2\rho(t)$ where $\rho(t)$ is the first-passage time distribution $\beta \delta c_0 = 0.2$ and various values of $\alpha$. (g) Optimal blind deposition intervals for fully reflecting flags (crosses) with corresponding MFPT (circles) and the approximations (\ref{eq:approx_Tbl_flags}) (dotted ad dashed lines). (h) Optimal values of $\alpha$ and $\delta c_0$ for the blind auto-chemotactic strategy as $\beta R_0/L$ varies. (k) Relative performance $\bar{T}_\mathrm{bl}/\bar{T}_\mathrm{inf} -1$ in the optimal strategies for the three models.  }
    \label{fig:bigfig}
\end{figure*}

{\it Informed strategies} --  For all three models, the optimal informed strategies are found analytically. The technical details of the derivations are available in a companion paper \cite{Meyer2026A}, we therefore only show the final results here. The optimal informed strategies in all three models result in a spatial concentration of the available resource as the budget decreases. The optimal resource allocation profiles are shown in figure \ref{fig:bigfig}(a-c).

Both models I and II can be investigated using Kramers formula for the MFPT of a one-dimensional diffusion process in a potential profile $V(x)$, namely $\bar{T} = D^{-1} \int_{0}^L dx \int_0^x dy e^{\beta (V(x)-V(y))}$\cite{Redner2001}. In model I, we seek a conservative force field, i.e. $F(x)=-V'(x)$ that minimizes the MFPT under the finite work budget constraint with no credit along the trajectory, that is $V(0)-V(L) = W$ and $\max_x V(x) - \min_x V(x) = W$. The optimal potential profile that satisfies this is found to be piecewise linear, and therefore the force profile piecewise constant, namely $F(x) = F^*$ for $x_- \leq x \leq x_+$ and $F(x)=0$ otherwise.
The region of the non-vanishing force $F^*$ is symmetric around the center of the domain, i.e. $x_\pm = L/2 \pm \Delta^*$. These optimal quantities, together with the corresponding optimal MFPT $\bar {T}_{\mathrm{inf}}^*$ read
\begin{equation}
\Delta^* =  \frac{\beta W L }{2+\beta W} , \ \ \  F^* =  \frac{2+\beta W}{\beta L} , \ \ \ 
 \bar {T}_{\mathrm{inf}}^* = \frac{\bar{2\mathcal{T}}_{0}}{2+\beta W}.
\end{equation}

In model II, informed protocols consist in predetermining the locations where flags will be deposited as the searcher hops over them for the first time. Since flags affect only backward jumps, while the probability of jumping from site $i$ to site $i+1$ remains independent of their presence, depositing a flag as the walker crosses a bond {\it for the first time} is equivalent to placing flags at predetermined positions beforehand and letting the searcher evolve in a static flag profile. In the continuous limit, it is therefore equivalent to a diffusive walk in a fixed piecewise constant potential made of $N$ discrete downward jumps of height $\beta^{-1}|\ln(1-p)|$. The MFPT is minimized when the flag locations $x_k$ form an arithmetic progression, $x_k=x_1+(k-1)\delta$, where $x_1$, $\delta$ and the corresponding MFPT, $\bar{T}_{\mathrm{inf}}^*$, are given by
\begin{equation}
x_1=\frac{L}{2+(N-1)p}, \ \ \delta= px_1, \ \ \bar{T}_{\mathrm{inf}}^* = \frac{(2-p)\bar{\mathcal{T}}_0}{2+(N-1)p}
\end{equation}
This profile is again symmetric around the midpoint of the interval $[0,L]$. In both models I and II, a low resource supply, i.e. low values of $W$ or $N$, requires to allocate the resource in a narrow region of space halfway through the search.

Model III differs markedly from the first two in that the informed protocol is not equivalent to a walk in a fixed potential. In fact, for each site $i$ the searcher deposits a predetermined amount $c^*_i$ whenever it reaches it for the first time.  Because the presence of cue impacts both the forward and backward motion, the jump probabilities as defined in Eq.~(\ref{eq:TSAW}) vary depending on whether the current site $i$ of the searcher is the furthest it has ever reached or it has previously been explored. The MFPT in the discrete setting can be derived analytically and its continuous limit follows as
\begin{align}
    \label{eq:kramers_autochemo}
    \bar{T}_{\mathrm{inf}} = &  \frac{1}{D}\int_{0}^L dx \int_{0}^x dy e^{\beta (c^*(x)-2c^*(y))} 
\end{align}
where $c^*(x)$ is the continuous version of the cue profile aimed to be generated. This expression resembles Kramers formula, but includes an additional prefactor in the exponential that accounts for the difference in transition probabilities between previously visited and unvisited sites. For an appropriately chosen cue field $c(x)$, this factor can substantially reduce the MFPT. The Euler-Lagrange equation aimed at minimizing Eq. (\ref{eq:kramers_autochemo}) under the finite resource constraint $\int_0^L dx c^*(x) = R_0$ and the positivity constraint $c^*(x)\geq 0$ can be solved   exactly. The optimal cue profile $c(x)$ is found to be a two-part piecewise function characterized by a threshold location $x_\circ$. For $x<x_\circ$, we have
\begin{align}
    \beta c^* (x) =& z_0 + 2 + \mathcal{W}_{-1}\left(\frac{(2z_0 + 3)e^{-(z_0+2)}x}{4(x_\circ - L)}- 2 e^{-2}\right)
\end{align}
where $\mathcal{W}_{-1}$ denotes the negative branch of the Lambert function, and $c^*(x)=0$ for $x\geq x_\circ$. The initial value $z_0 = \beta c^*(0)$ is defined as the solution of
\begin{gather}
    \label{eq:z0}
     z_0 - \frac{3}{2}  + \frac{ (2z_0 + 3)^2 e^{-z_0}}{16 - (4z_0 + 10)e^{-z_0}} = \frac{\beta R_0}{L}
\end{gather}
and the threshold position $x_\circ$ is obtained in terms of $z_0$, namely $x_\circ = (4z_0 L - 2\beta R_0)/(2z_0+3)$.  In the limit $\beta R_0 \ll L$, we have $z_0 \simeq \sqrt{3\beta R_0/2L}$, $x_\circ \simeq L\sqrt{8\beta R_0 /3L}$, and $\bar{T}_{\mathrm{inf}} \simeq \bar{\mathcal{T}}_0 \left[ 1 - 2\beta R_0/L \right]$ at lowest order in $\beta R_0/L$: the cue profiles are initially steep and almost linear until $x_\circ$, which remains small compared to $L$. On the other end in the limit $\beta R_0\gg L$, we obtain $z_0 \simeq \beta R_0 /L - \frac{3}{2}$, $x_\circ \simeq L - e^{-\beta R_0 /L} \beta R_0/4$, and $\bar{T}_{\mathrm{inf}} \simeq 3\bar{\mathcal{T}}_0e^{-\beta R_0 /L}$. This corresponds to almost rectangular profiles for which the MFPT decreases exponentially fast as $R_0\to \infty$. Similarly to models I and II, the informed protocol of model III concentrates resource usage in a narrow region of space when the available supply is limited. However,it does so in the early stages of the walk, as a consequence of the lack of directional information carried by the cue. Since the only bias towards the target available to the searcher arises from the reflecting boundary at $x=0$, the strong initial deposition allows to exploit it for as long as possible, enhancing motion towards the target from the start of the search. As $R_0$ increases, large deposition rates become sustainable for longer times, thus increasing the gradient at the front the profile and eventually leading to near-ballistic motion. This results in an optimal MFPT substantially lower than those obtained with external pulling and directional flags. \\

{\it Blind strategies} -- If the searcher cannot condition the allocation of resource on its current position and past trajectories, it must therefore decide of it based on a predetermined temporal protocol. Mean resource usage profiles are shown in figure \ref{fig:bigfig}(d-f).

In model I, the searcher starts applying a constant force $F$ at time $\tau$ and stops whenever the work budget $W$ is fully exhausted, unless it reaches the target before. The MFPT for this process can be found analytically and requires to distinguish between two cases, depending whether the distance $W/F$ over the which the budget becomes exhausted exceeds $L$. The MFPT is minimized for $W/F<L$ because it allows the budget to be fully used, but the optimal switching time $\tau$ is very short relative to the diffusion time scale. Although the exact expression for this time is not accessible, imposing $\tau=0$ and optimizing the MFPT with respect to $F$ approaches the true minimum remarkably closely. In such a case, the optimal force is simply $F=W/L$, i.e. this near-optimal strategy consists in switching on the force from the start of the process with a force that exhausts the work budget precisely when the target is reached. The corresponding optimal MFPT is then $\bar{T}_\mathrm{bl} = 2\mathcal{\bar{T}}_0 \left[e^{-\beta W} - 1 + \beta W\right]/(\beta W)^2$. This strategy noticeably differs from the informed strategies in which the application of the force is concentrated in a narrow region of space.  

This observation remains true for model II, for which blind strategies consist in predetermining a set of times $\tau_k$ at which the $k$-th flag will be deposited. The case of totally reflecting flags $p=1$ is particularly insightful as flags effectively act as walls which simplifies the mathematical structure of the problem into a set of convolution products and allows for analytical predictions. By first assuming that the time interval between two depositions is constant and noted $\Delta$, and then assuming that all trajectories exhaust the flag supply, which is asymptotically true as $\Delta$ is small, we can approximate the MFPT as 
\begin{align}
    \bar{T}_{\mathrm{bl}} \simeq& \mathcal{T}_0  - 2N\sqrt{\frac{L^2 \Delta}{D \pi }} + 2N\left(1 + \frac{N-1}{\pi}\right) \Delta, 
\end{align}
the next term being of order $(D\Delta/L^2)^{3/2}$. This approximation yields a minimum $\bar{T}_{\mathrm{bl}}^*$ found for a deposition interval $\Delta^*$ reading 
\begin{align}
    \label{eq:approx_Tbl_flags}
    \Delta^* =\frac{\pi\bar{\mathcal{T}}_0 }{2(N-1+\pi)^2}, \qquad \bar{T}_{\mathrm{bl}}^* =&  \frac{(\pi - 1)\bar{\mathcal{T}}_0 }{N-1+\pi}
\end{align}
These approximations are actually very close to the actual optimal strategy identified by gradient descent algorithm, which shows that the deposition intervals $\Delta_k = \tau_{k} - \tau_{k-1}$ are almost constant and well approximated by Eq.~(\ref{eq:approx_Tbl_flags}) -- see figure \ref{fig:bigfig}g. The accuracy of this approximation decreases with the number of flags $N$ since assuming the flag supply to be exhausted in all trajectories becomes increasingly inaccurate. For the partially reflecting flags with $p<1$, Brownian dynamics simulations performed with a wide range of randomly chosen introduction times show that the MFPT is minimized when the introduction times are evenly separated by a constant deposition interval $\Delta$, within the uncertainty due to finite sampling. Similarly to model I, the blind optimal strategy therefore requires to use resources at a steady pace in model II, unlike the informed one. 

While models I and II still perform reasonably well in the optimal blind protocols, the blind auto-chemotactic process suffers from a well-known effect encountered in self-repelling walks, namely self-trapping \cite{Grassberger2017}. The cue profile generated by the walker along its path may be non-monotonic as the walker's trajectory is not ballistic, which forms barriers. While an infinite cue supply allows to overcome these barriers by filling the valleys with more cue, this can only be done until the cue supply is too low in the finite case. Low enough barriers can still be overcome in a reasonable time once the supply is exhausted, but too high barriers lead to exceptionally long search times, producing a first-passage time distribution with a slowly decaying algebraic tail that may result in a divergent mean. Considering deposition protocols in which the amount deposited is adjusted based on the remaining supply, i.e. $\delta c_n = \delta c_0 \left(R_n/R_0 \right)^\alpha$ where $\delta c_n$ is the increment in the cue profile deposited at the searcher's site at time step $n$, we find that the MFPT is minimized for values of $\alpha$ and $\delta c_0$ that increases with $R_0$. The deposition rate should therefore not be constant but instead faster in the early stages of the process. Still, the trajectory-averaged cue profile slowly decreases and reaches supply exhaustion at the target location. This mitigating strategy is however not extremely efficient to reduce the MFPT, which optimally decreases mildly with $R_0$. The blind strategies are therefore mostly driven by a risk-averse criterion: attempting to accelerate faster trajectories, thereby shifting the first-passage time distribution towards shorter times, inevitably comes at the cost of significantly heavier long-time tails. The priority should therefore be to suppress these tails before seeking to further accelerate the search. \\

{\it Relative performance} -- While both informed and blind strategies allow to reduce the MFPT if properly optimized, their relative performance significantly differ from one model to the other.  In the external pulling process, the ratio $\bar{T}_\mathrm{bl}/\bar{T}_\mathrm{inf}$ shows a clear maximum at an intermediate value of the work budget. It then decays as  $W^{-1}$, i.e. informed and blind strategies are as efficient in the large budget limit. 
For the directional flags, the ratio also initially increases as the number of flags $N$ increases but then settles to a plateau value. Finally, the auto-chemotactic process shows the most dramatic effect as the informed strategy leads to an optimal MFPT that decays exponentially with the resource supply $R_0$, unlike the much less efficient blind strategy. As a result the ratio $\bar{T}_\mathrm{bl}/\bar{T}_\mathrm{inf}$ grows monotonically with the supply size $R_0$ and diverges as $R_0\to \infty$. \\

\textit{Outlook} -- Our results highlight key features of resource-limited search strategies. Informed strategies concentrate resource use when the supply is low, exploiting knowledge of the current progress of the search process to ensure depletion before reaching the target. The optimal deposition pattern depends on the cues: directional cues allow resource use to be concentrated at the most advantageous point, whereas symmetric cues rely on other sources of positional information, such as the initial condition of the walk. Continuously divisible cues, as in Model III, offer further flexibility to fine-tune deposition and the resulting effective forces, leading to faster search times. 
	
	Blind strategies, by contrast, favor a more uniform resource usage due to stochastic fluctuations, which prevent reliable synchronization of resource depletion with target arrival. Delaying resource use risks reaching the target before any resources are deployed. Symmetric cues have the additional drawback that they can drive the searcher away from the target or trap it for extended periods. A constant usage rate therefore emerges as optimal.
	
	These findings extend to targets randomly distributed, in particular for a homogeneously distributed target over the interval $[0,L]$ \cite{Meyer2026A}. Informed strategies still concentrate resource use in increasingly narrow spatial regions as the supply decreases, with the deposition region shifted towards smaller $x$. Blind strategies retain a constant optimal usage rate, slightly higher than in the fixed-target case to account for targets near the beginning of the domain. The scaling of all relevant quantities with supply size remains unchanged, demonstrating the robustness of our results across different search settings.  \\

{\it Acknowledgements} -- This work was supported by the Medical Research Council (Grant No. MR/Y003845/1) and the Engineering and Physical Sciences Research Council (Grant No. EP/ V031201/1)

%\bibliographystyle{apsrev4-2}
%\bibliography{references}

%apsrev4-2.bst 2019-01-14 (MD) hand-edited version of apsrev4-1.bst
%Control: key (0)
%Control: author (72) initials jnrlst
%Control: editor formatted (1) identically to author
%Control: production of article title (-1) disabled
%Control: page (0) single
%Control: year (1) truncated
%Control: production of eprint (0) enabled
%

\end{document}